\documentclass[aps,prd,a4paper,twocolumn,amsmath,showpacs,superscriptaddress,nofootinbib,preprintnumbers]{revtex4-1}
\usepackage{graphicx,ulem}
\usepackage{longtable}
\usepackage{float}
\usepackage{dcolumn}
\usepackage{graphics,epsfig}
\usepackage{amsmath,amssymb,latexsym,mathrsfs}
\usepackage{bm}
\usepackage{color}
\usepackage{color}
\usepackage{subfigure}
\usepackage{hyperref}
\usepackage{wasysym}\usepackage{array}

\newcommand{\clkg}{$C_\ell^{\rm \kappa g}\,$}
\newcommand{\clkgb}{$\boldsymbol{C_\ell^{\rm \kappa g}\,}$}

\newcommand{\mnu}{M_{\nu}}
\usepackage{titlesec}
\titleformat{\paragraph}
{\normalfont\normalsize\bfseries}{\theparagraph}{1em}{}
\titlespacing*{\paragraph}
{0pt}{3.25ex plus 1ex minus .2ex}{1.5ex plus .2ex}

\begin{document}
\title{Scale-dependent galaxy bias, CMB lensing-galaxy cross-correlation, and neutrino masses}

\author{Elena Giusarma}
\email{egiusarma@lbl.gov}
\affiliation{Lawrence Berkeley National Laboratory (LBNL), Physics Division, Berkeley, CA 94720-8153, USA}
\affiliation{Berkeley Center for Cosmological Physics, University of California, Berkeley, CA 94720, USA}
\affiliation{McWilliams Center for Cosmology, Department of Physics, Carnegie Mellon University, Pittsburgh, PA 15213, USA}

\author{Sunny Vagnozzi}
\email{sunny.vagnozzi@fysik.su.se}
\affiliation{The Oskar Klein Centre for Cosmoparticle Physics, Department of Physics, Stockholm University, \\ AlbaNova Universitetscentrum, Roslagstullbacken 21A, SE-106 91 Stockholm, Sweden}
\affiliation{The Nordic Institute for Theoretical Physics (NORDITA), Roslagstullsbacken 23, SE-106 91 Stockholm, Sweden}

\author{Shirley Ho}
\affiliation{Lawrence Berkeley National Laboratory (LBNL), Physics Division, Berkeley, CA 94720-8153, USA}
\affiliation{Berkeley Center for Cosmological Physics, University of California, Berkeley, CA 94720, USA}
\affiliation{McWilliams Center for Cosmology, Department of Physics, Carnegie Mellon University, Pittsburgh, PA 15213, USA}
\affiliation{Center for Computational Astrophysics, Flatiron Institute, 162 5th Avenue, New York, NY 10010, USA}

\author{Simone Ferraro}
\affiliation{Berkeley Center for Cosmological Physics, University of California, Berkeley, CA 94720, USA}
\affiliation{Miller Institute for Basic Research in Science, University of California, Berkeley, CA, 94720 USA}
\affiliation{Lawrence Berkeley National Laboratory (LBNL), Physics Division, Berkeley, CA 94720-8153, USA}

\author{Katherine Freese}
\affiliation{The Oskar Klein Centre for Cosmoparticle Physics, Department of Physics, Stockholm University, \\ AlbaNova Universitetscentrum, Roslagstullbacken 21A, SE-106 91 Stockholm, Sweden}
\affiliation{The Nordic Institute for Theoretical Physics (NORDITA), Roslagstullsbacken 23, SE-106 91 Stockholm, Sweden}
\affiliation{Leinweber Center for Theoretical Physics, Department of Physics, University of Michigan, Ann Arbor, MI 48109, USA}

\author{Rocky Kamen-Rubio}
\affiliation{Lawrence Berkeley National Laboratory (LBNL), Physics Division, Berkeley, CA 94720-8153, USA}
\affiliation{Department of Physics, University of California, Berkeley, CA, 94720 USA}

\author{Kam-Biu Luk}
\affiliation{Lawrence Berkeley National Laboratory (LBNL), Physics Division, Berkeley, CA 94720-8153, USA}
\affiliation{Department of Physics, University of California, Berkeley, CA, 94720 USA}

\date{\today}

\begin{abstract}
One of the most powerful cosmological datasets when it comes to constraining neutrino masses is represented by galaxy power spectrum measurements, $P_{gg}(k)$. The constraining power of $P_{gg}(k)$ is however severely limited by uncertainties in the modeling of the scale-dependent galaxy bias $b(k)$. In this work we present a new proof-of-principle for a method to constrain $b(k)$ by using the cross-correlation between the Cosmic Microwave Background (CMB) lensing signal and galaxy maps (\clkg) using a simple but theoretically well-motivated parametrization for $b(k)$. We apply the method using \clkg measured by cross-correlating \textit{Planck} lensing maps and the Baryon Oscillation Spectroscopic Survey (BOSS) Data Release 11 (DR11) CMASS galaxy sample, and $P_{gg}(k)$ measured from the BOSS DR12 CMASS sample. We detect a non-zero scale-dependence at moderate significance, which suggests that a proper modeling of $b(k)$ is necessary in order to reduce the impact of non-linearities and minimize the corresponding systematics. The accomplished increase in constraining power of $P_{gg}(k)$ is demonstrated by determining a 95\% confidence level upper bound on the sum of the three active neutrino masses $M_{\nu}$ of $M_{\nu}<0.19\, {\rm eV}$. This limit represents a significant improvement over previous bounds with comparable datasets. Our method will prove especially powerful and important as future large-scale structure surveys will overlap more significantly with the CMB lensing kernel providing a large cross-correlation signal.
\end{abstract}

\pacs{}
\maketitle

\section{Introduction}\label{sec:intro}

Galaxies, due to complexities inherent to their formation and evolution, are biased tracers of the underlying matter distribution. In other words, the galaxy power spectrum measured from redshift surveys, $P_{gg}(k,z)$, is related to the underlying matter power spectrum $P(k,z)$ (which cannot be directly measured, but represents the true source of cosmological information) through a factor $b$ known as \textit{bias}~\cite{Desjacques:2016bnm}:
\begin{eqnarray}
P_{gg}(k,z) \approx b_{\rm auto}^2P(k,z) \, ,
\label{galaxy}
\end{eqnarray}
The subscript ``auto'' refers to the fact that $P_{gg}(k,z)$ is an auto-correlation quantity, since it corresponds to the Fourier transform of the 2-point auto-correlation function of the galaxy overdensity field, $\xi(r)$.

Galaxy bias also enters in cross-correlation quantities, such as the matter-galaxy cross-power spectrum $P_{mg}(k,z)$. This quantity is given by the Fourier transform of the 2-point cross-correlation function between the matter (dark matter plus baryons) and galaxy overdensity fields, $\xi^{mg}(r)$. However, the bias appearing in $P_{mg}(k,z)$ differs from that of Eq.~(\ref{galaxy}):
\begin{eqnarray}
P_{mg} (k,z) \approx b_{\rm cross}P(k,z) \, .
\label{pmg}
\end{eqnarray}
The difference between $b_{\rm auto}$ and $b_{\rm cross}$, explained more in detail in Sec.~\ref{sec:theo}, is expected based on results of N-body simulations~\cite{Villaescusa-Navarro:2013pva,Okumura:2012xh,Hand:2017ilm,Modi:2017wds,Vlah:2013lia}, as well as theoretical arguments.

Heretofore, the bias has often been modeled as a scale-independent quantity in cross-correlation analysis~\cite{Giusarma:2013pmn,Cuesta:2015iho, Giusarma:2016phn,Vagnozzi:2017ovm}. However, this approach is truly reliable only on large, linear scales ($k<k_{\max}=0.15\, h{\rm Mpc}^{-1}$ today and $k<k_{\max}=0.2\, h{\rm Mpc}^{-1}$ at a redshift of about $0.5$)~\cite{Desjacques:2016bnm}, therefore preventing one from fully retrieving information on cosmological parameters. The simplest and best-motivated forms of the scale-dependent biases read~\cite{Desjacques:2016bnm,Sheth:1999mn,Seljak:2000jg,Schulz:2005kj,Smith:2006ne,Manera:2009zu,Desjacques:2010gz,Musso:2012ch,
Paranjape:2012ks,Schmidt:2012ys,Verde:2014nwa,Senatore:2014eva,Castorina:2016pqq,Modi:2016dah}:
\begin{eqnarray}
b_{\rm cross}(k) = a+c k^2 \, ,
\label{biascross} \\
b_{\rm auto}(k) = a+d k^2 \, ,
\label{biasauto}
\end{eqnarray}
\\
where $a$, $c$ and $d$ are three free parameters describing the scale-dependent bias. It is worth remarking that, while various phenomenological expressions for $b(k)$ abound in the literature (although see~\cite{Smith:2006ne} for earlier criticisms related to phenomenological parametrizations), the expression we use is extremely well motivated on both theory and simulations grounds. As a token of the robustness of this model, it is remarkable that at least three well-known but distinct theoretical approaches to the study of galaxy bias (peaks theory~\cite{Desjacques:2010gz}, the excursion set approach~\cite{Musso:2012ch}, and the effective field theory of large-scale structure~\cite{Senatore:2014eva}) predict \textit{exactly the same} functional form for $b(k)$ in the mildly non-linear regime that we are interested in, with results from simulations agreeing with these findings (see Appendix for further discussions). In fact, in Fourier space, the lowest-order correction to a constant bias one can expect on general grounds, based on the sole assumption of isotropy, is a $k^2$ correction (a correction linear in $k$ would instead not respect isotropy).

Our goal is to provide a proof-of-principle for a correct and simple treatment enabling the retrieval of information on $b_{\rm auto}$ and $b_{\rm cross}$, in order to more robustly extract information from galaxy redshift surveys. To this end we require, in addition to galaxy power spectrum data [sensitive to $b_{\rm auto}$, Eq.~(\ref{galaxy})], measurements sensitive to the matter-galaxy cross-spectrum $P_{mg}(k)$ [containing information on $b_{\rm cross}$, Eq.~(\ref{pmg})]. Since the matter distribution is responsible for the gravitational lensing of CMB photons, we expect the cross-correlation between CMB lensing and galaxy overdensity maps, \clkg, to carry information on $P_{mg}(k)$ and hence on $b_{\rm cross}(k)$. Here $\kappa$ denotes the CMB lensing convergence.~\footnote{A CMB photon coming from a direction $\boldsymbol{\hat{n}}$ on the sky is deflected due to lensing by an angle $d(\boldsymbol{\hat{n}})=\boldsymbol{\nabla}\phi(\boldsymbol{\hat{n}})$, where $\phi(\boldsymbol{\hat{n}})$ is the lensing potential. The lensing convergence is then given by $\kappa(\boldsymbol{\hat{n}}) \equiv -\frac{1}{2}\boldsymbol{\nabla}^2\phi(\boldsymbol{\hat{n}})$.} The information one can extract on $b_{\rm cross}(k)$ (and therefore on $a$) is put to best use when combining \clkg measurements with galaxy power spectrum data $P_{gg}(k)$. The reason is that an improved determination of $b_{\rm auto}(k)$ (through the improved constraints on $a$) significantly bolsters the constraining power of the galaxy power spectrum. This improved determination is especially important for the estimation of cosmological parameters affecting the growth of structure, such as massive neutrinos.

Previous works have suggested combining lensing and clustering (power spectrum) measurements~\cite{Giannantonio:2015ahz, Joudaki:2017zdt} or adopting a scale-dependent galaxy bias parametrization~\cite{Pen:2004rm, More:2014uva, Amendola:2015pha, Beutler:2016zat, Beutler:2016arn}. In this paper, it is the \textit{first time} that:
\begin{itemize}
\item \clkg and $P_{gg}(k)$ measurements are combined, interpreted and analyzed in light of the simple but well-motivated~\cite{Desjacques:2016bnm,Sheth:1999mn,Seljak:2000jg,Schulz:2005kj,Smith:2006ne,Manera:2009zu,Musso:2012ch,
Paranjape:2012ks} scale-dependent biases models given by Eqs.~(\ref{biascross},\ref{biasauto}).
\item The achieved increase in constraining power of $P_{gg}(k)$ is used to extract tighter and more robust limits on the sum of the neutrino masses $M_{\nu}$. We show that our limits on $M_{\nu}$ are substantially strengthened when compared to previous results obtained through a scale-independent treatment of the bias~\cite{Cuesta:2015iho, Giusarma:2016phn, Vagnozzi:2017ovm}.
\end{itemize}
This work should be seen as a proof-of-principle of our methodology, rather than a fully fledged analysis. There are several aspects of our method and analysis that deserve a more in-depth investigation, as we shall discuss later in our paper: we plan to return to these issues in future work.

\section{Theory}\label{sec:theo}

To obtain information on cosmological parameters from \clkg, one must be able to model the theoretical prediction for \clkg given a set of cosmological parameters. Within a $\Lambda$CDM framework and adopting the Limber approximation~\cite{Limber:1954zz,LoVerde:2008re}, \clkg reads:
\begin{eqnarray}
\hskip -1 cm C_\ell^{\kappa g} = \int_{z_0}^{z_1} dz\frac{H(z)}{\chi^{2}(z)}W^{\kappa}(z)f_g(z)P_{mg}\left(k=\frac{\ell}{\chi(z)},z\right) \, .
\label{clkg}
\end{eqnarray}
The theoretical matter-galaxy cross-power spectrum $P_{mg}$ appearing on the right-hand-side of Eq.~(\ref{clkg}) is modeled following Eq.~(\ref{pmg}), with the theoretical $b_{\rm cross}(k)$ given by Eq.~(\ref{biascross}) and determined by the choice of parameters $a$ and $c$ in the MCMC analysis, while the theoretical non-linear matter power spectrum $P(k,z)$ is computed using the Boltzmann solver \texttt{CAMB}~\cite{Lewis:1999bs} and \texttt{Halofit}~\cite{Bird:2011rb,Takahashi:2012em} starting from the given cosmological parameters. Furthermore, $\chi(z)$ is the comoving distance to redshift $z$, $f_g(z)$ is the redshift distribution of the galaxy sample, $H(z)$ is the Hubble parameter, and $W^{\kappa}(z)$ is the CMB lensing convergence kernel~\cite{Peiris:2000kb,Hirata:2008cb,Bleem:2012gm,Sherwin:2012mr,Vallinotto:2013eva,Pearson:2013iha,
Bianchini:2014dla,Giannantonio:2015ahz,Pullen:2015vtb,Bianchini:2015yly,Singh:2016xey,Prat:2016xor,Bianchini:2018mwv}:
\begin{eqnarray}
W^{\kappa}(z)= \frac{3\Omega_{m,0}}{2c}\frac{H_0^2}{H(z)}(1+z)\chi(z)\frac{\chi(z_{_{\rm CMB}})-\chi(z)}{\chi(z_{_{\rm CMB}})} \, ,
\end{eqnarray}
where $H_0$ and $\Omega_{m,0}$ denote the Hubble parameter and matter density at present time. Comparing the theoretical prediction for \clkg [right-hand side of Eq.~(\ref{clkg})] to its measured value through the likelihood function allows us to derive constraints on $b_{\rm cross}(k)$. In Eq.~(\ref{clkg}) we have chosen for simplicity not to include the contribution of redshift-space distortions, as well as the contribution of lensing to the observed galaxy clustering. The former is negligible on the scales of interest, whereas~\cite{Dizgah:2016bgm} showed that neglecting the latter at $z=0.57$ induces a relative error of less than $5\%$ in $C_{\ell}^{\kappa g}$, which is well below the current error budget in the measured $C_{\ell}^{\kappa g}$.  \\

From peaks theory~\cite{Bardeen:1985tr}, as well as on more general grounds, one expectes differences between $b_{\rm cross}(k)$ and $b_{\rm auto}(k)$ [Eqs.~(\ref{biascross},\ref{biasauto})]. To some extent, these differences are partly attributable to stochasticity~\cite{Desjacques:2016bnm,Sheth:1999mn,Seljak:2000jg,Schulz:2005kj,Smith:2006ne,Manera:2009zu,Musso:2012ch,
Paranjape:2012ks,Schmidt:2012ys,Verde:2014nwa,Castorina:2016pqq,Modi:2016dah} (see also Figs.~1 and 2 of~\cite{Villaescusa-Navarro:2013pva}). The stochastic component, which is expected to be scale-dependent and hence more complex than a simple white shot-noise component~\cite{Baldauf:2013hka}, originates from the discrete nature of galaxies as tracers of the density field, as well as the non-Poissonian behavior of satellite galaxies whose spatial distribution does not follow that of the dark matter in halos~\cite{Dvornik:2018frx}. Auto-power spectra measurements therefore include a stochastic component, whereas cross-power spectra measurements are substantially less sensitive to the stochastic component. We take into account this difference by considering two separate parameterizations for $b_{\rm cross}$ and $b_{\rm auto}$ as per Eqs.~(\ref{biascross},\ref{biasauto}).~\footnote{Note that a relation between the bias parameters $c$ [Eq.~(\ref{biascross})] and $d$ [Eq.~(\ref{biasauto})] is still not present in the literature.} Eq.~(\ref{biascross}) and Eq.~(\ref{biasauto}) are used to model the theoretical values of \clkg [Eq.~(\ref{clkg})] and $P_{gg}(k)$ [Eq.~(\ref{galaxy})] respectively when comparing them to their measured values in the likelihood function, allowing us to derive constraints on the bias parameters $a$, $c$, and $d$.

Note that, on simulations grounds, $b_{\rm cross}$ is typically expected to increase with increasing $k$ (\textit{i.e.} $db_{\rm cross}/dk>0$), whereas the opposite behaviour is expected for $b_{\rm auto}$ (\textit{i.e.} $db_{\rm auto}/dk<0$). To see this behaviour in simulations of luminous red galaxies (LRGs, which we will use in our work) at $z=0.5$, see the light blue short-dashed and long-dashed curves in the second panel from the left of the upper row of Fig.~2 in~\cite{Okumura:2012xh}. This behaviour is even more enhanced for more massive and hence more biased galaxies, see the purple and dark blue curves in the same figure.~\footnote{For LRGs, $b_{\rm cross}$ and $b_{\rm auto}$ appear to be nearly equal up to $k \sim 0.2\, h{\rm Mpc}^{-1}$, suggesting that in principle we could have taken $c=d$. However, in order to be conservative we have decided to allow the two scale-dependent factors to be independent. In fact, as we shall see later, data ends up detecting differences between $c$ and $d$.} On theoretical grounds, such a behaviour is not unexpected. Concerning $b_{\rm cross}$, it is known that on small scales the matter-galaxy 2-point correlation function $\xi^{mg}(r)$ traces the halo density profile $\rho(r)$ (see e.g. Fig.~1 in~\cite{Hayashi:2007uk}) and hence rises steeply. One therefore expects $b_{\rm cross}$ to rise on small scales (large $k$), as seen in simulations. Turning to auto-correlation measurements instead, halos are extended objects and therefore the distance between halos cannot be less than the sum of their radii: this effect of \textit{halo exclusion} is translated into the fact that, on small scales, the galaxy 2-point correlation function $\xi(r) \to -1$~\cite{CasasMiranda:2001ym,Smith:2006ne,Baldauf:2013hka}. Therefore, one expects $b_{\rm auto}$ to drop on small scales (large $k$), again in agreement with what is observed in simulations. This justifies our choice of treating $b_{\rm cross}$ and $b_{\rm auto}$ separately, albeit using the same functional form for both, which is justified on both theory and simulations grounds.

\section{Datasets and methodology}\label{sec:data}

The baseline dataset we consider consists of measurements of the CMB temperature, polarization, and cross-correlation spectra from the Planck 2015 data release~\cite{Ade:2015xua, Adam:2015rua, Aghanim:2015xee}. We combine the high-$\ell$ and low-$\ell$ temperature likelihoods, as well as the low-$\ell$ polarization likelihood. This dataset combination is referred to as \textbf{\textit{CMB}}.

In addition, we also include the galaxy power spectrum data from the BOSS DR12 CMASS sample~\cite{Alam:2016hwk,Gil-Marin:2015sqa}. We denote this dataset by $\boldsymbol{P_{gg}(k)}$. The measured galaxy power spectrum is compared to the theoretical value through the likelihood function, where the theoretical galaxy power spectrum $P_{gg}^{\rm th}(k,z)$ is modeled as follows:
\begin{eqnarray}
\hskip -1 cm P_{gg}^{\rm th}(k,z) = b_{\rm auto}^2(k) \left ( 1+\frac{2}{3}\beta + \frac{1}{5}\beta^2 \right )P_{\rm HF\nu}(k,z)+P^{\rm s} \, .
\label{Pgg_rsd}
\end{eqnarray}
In Eq.~(\ref{Pgg_rsd}), $\beta = \Omega_m(z_{\text{eff}})^{0.545}/b_{\rm auto}(k)$ parametrizes the amplitude of redshift-space distortions at the effective redshift $z_{\text{eff}} = 0.57$ determined by the BOSS collaboration~\cite{Alam:2016hwk,Gil-Marin:2015sqa}, and $b_{\rm auto}(k)$ is given in Eq.~(\ref{biasauto}).~\footnote{We also verified that if we consider a linear redshift distortion parameter, $\beta = \Omega_m(z_{\text{eff}})^{0.545}/a$, this choice has no effects on our results.} $P_{\rm HF\nu}(k,z)$ is the theoretical non-linear matter power spectrum computed using \texttt{Halofit}~\cite{Bird:2011rb, Takahashi:2012em}. Notice that we do not model non-linear redshift-space distortions in Eq.~(\ref{Pgg_rsd}) because their contribution on the scales of interest ($k<0.2\,h{\rm Mpc}^{-1}$) is small (see e.g Figure 5 of~\cite{Okumura:2015fga}). Finally, $P^{\rm s}$ is a nuisance parameter taking into account residual shot-noise contribution due to the discrete nature of galaxies. We consider the same wavenumber range used in~\cite{Vagnozzi:2017ovm}, $0.03\, h{\rm Mpc}^{-1} < k < 0.2\, h{\rm Mpc}^{-1}$, in order to avoid the use of non-linear scales, which would require a more sophisticated bias model beyond the relatively simple one we are using. In future work we will explore how a more sophisticated bias model can allow us to push to more non-linear scales.

In addition to the CMB and galaxy power spectrum data, we consider the cross-correlation, measured by Pullen \textit{et al.}~\cite{Pullen:2015vtb},  between CMB lensing convergence maps from the Planck 2015 data release~\cite{Ade:2015zua} and galaxy overdensity maps from the DR11 CMASS sample~\cite{Anderson:2013zyy}. We refer to this dataset as \clkgb. Following~\cite{Pullen:2015vtb}, we limit our use of the measurements of \clkg from $\ell=130$ to $\ell=950$, thus removing the points in the low-$\ell$ range. The choice is dictated by the observed discrepancy between measurements of $P_{gg}(k)$ in the North and South Galactic caps~\cite{Ross:2012qm}, as well as possible contamination from the thermal Sunyaev-Zel'dovich (SZ) effect or other unknown systematics on large angular scales, to be discussed briefly later. This observation suggests that large-scale clustering measurements could be affected by systematics (see also~\cite{Hahn:2016kiy}).

It is worth pointing out that \clkg measurements are extremely valuable due to their ability of breaking the degeneracy between $a$ and $\sigma_8$. While $P_{gg}$ is sensitive to the quantity $a^2\sigma_8^2$, \clkg is instead sensitive to the combination $a\sigma_8^2$. The combination of \clkg and $P_{gg}$ is thus capable of breaking the degeneracy between the parameters $a$ and $\sigma_8$.

We assume the standard six-parameter $\Lambda$CDM cosmological model, complemented by four parameters describing the scale-dependent bias ($a$, $c$, and $d$) and the sum of the three active neutrino masses $M_{\nu}$. For $M_{\nu}$ we adopt the currently sufficiently precise assumption of a degenerate mass spectrum~\cite{Lesgourgues:2004ps,DeBernardis:2009di,Wagner:2012sw,Gerbino:2016sgw,Archidiacono:2016lnv,Lattanzi:2017ubx, Lesgourgues:2006nd}. We do not model the modification to the scale-dependent bias induced by massive neutrinos~\cite{Lesgourgues:2009am,Saito:2009ah,Shoji:2010hm,Ichiki:2011ue,Dupuy:2013jaa,
Biagetti:2014pha,LoVerde:2014pxa,LoVerde:2014rxa,Blas:2014hya,Fuhrer:2014zka,
Dupuy:2015ega,Archidiacono:2015ota,Levi:2016tlf,Raccanelli:2017kht,Senatore:2017hyk,Brandbyge:2008rv,Viel:2010bn,
Brandbyge:2010ge,Agarwal:2010mt,Marulli:2011he,AliHaimoud:2012vj,Villaescusa-Navarro:2013pva,
Castorina:2013wga,Costanzi:2013bha,Baldi:2013iza,Massara:2014kba,Castorina:2015bma,Carbone:2016nzj,
Banerjee:2016zaa,Rizzo:2016mdr,Villaescusa-Navarro:2017mfx,Chiang:2017vuk,Munoz:2018ajr,Vagnozzi:2018pwo}, as~\cite{Raccanelli:2017kht,Vagnozzi:2018pwo} found that this effect is negligible given the sensitivity of current data.

We sample the posterior distributions of the cosmological parameters using the publicly available MCMC sampler \texttt{CosmoMC}~\cite{Lewis:2002ah,Lewis:2013hha}. We assume a Gaussian likelihood for \clkg, with covariance matrix estimated by jackknife resampling~\cite{Pullen:2015vtb}. The theoretical values of $P_{gg}$ and $C_{\ell}^{\kappa g}$ are convolved with the respective window functions, which take into account the finite geometry of the surveys, before being compared to their measured values in the likelihood function.

Unless otherwise specified, a uniform prior is assumed for all cosmological parameters. We allow $M_{\nu}$ to be as small as $0\, {\rm eV}$, ignoring prior information from oscillation experiments, which set a lower limit of $0.06\, {\rm eV}$~\cite{Esteban:2016qun,Capozzi:2017ipn,deSalas:2017kay}.~\footnote{This choice for the lower limit of the $M_{\nu}$ prior can also be viewed as a phenomenological proxy for models where the neutrino energy density can be smaller than the one predicted in $\Lambda$CDM, if not vanishing, see e.g.~\cite{Beacom:2004yd}.} For completeness we also report constraints on $M_{\nu}$ when this lower limit is imposed. For $a$ we impose a uniform prior in the range between 0 and 5, while for $c$ and $d$ we adopt a uniform prior between -50 and 10 (in units of $h^{-2}\,{\rm Mpc}^2$). The choice for the lower ranges of $c$ and $d$ is dictated by N-body simulations~\cite{Villaescusa-Navarro:2013pva,Vlah:2013lia}. These prior ranges are large enough to not cut the respective posterior distributions where these are significantly different from zero: in other words, the data really will be deciding the preferred ranges of $c$ and $d$, and not the priors.

\section{Results}\label{sec:results}

Table~\ref{tab:tab1} shows the constraints we obtain on $a$, $c$, $d$, and $M_{\nu}$, for various datasets combinations. We begin by considering the \textbf{\textit{CMB}} CMB-only dataset, and find $M_{\nu}<0.72\, {\rm eV}$ at $95\%$~C.L.~\cite{Ade:2015xua}.

The addition of \clkgb (second and third rows of Table~\ref{tab:tab1}) allows us to constrain $a$ and $c$. We find $a \simeq 1.5 \pm 0.2$ at $1\sigma$, a value which is low when compared to the expectation from simulations for this galaxy sample ($a \approx 2$~\cite{Alam:2016hwk,Gil-Marin:2015sqa}), although compatible at $\approx 2.5\sigma$. We attribute this low value to a deficit of large-scale power observed in several measurements of \clkg~\cite{Kuntz:2015wza,Giannantonio:2015ahz}, including ours. Explanations range from systematics introduced in the \textit{Planck} 2015 lensing maps~\cite{Omori:2015qda,Liu:2015xfa,Kuntz:2015wza} to contamination from thermal SZ~\cite{vanEngelen:2013rla}.

The observed deficit in power also affects the bounds on $c$, because $a$, $c$, and $M_{\nu}$ are mutually degenerate when considering \clkg measurements only. The reason is that a decrease in $a$ can be compensated on small scales by increasing $c$. An increase in $c$ increases power on small scales: this can be compensated by increasing $M_{\nu}$ in order to damp small-scale power.

The fourth and fifth rows of Table~\ref{tab:tab1} report the bounds obtained from the \textbf{\textbf{\textit{CMB}}}+$\boldsymbol{P_{gg}(k)}$ dataset. In this case $a$ and $d$ do not show a strong degeneracy. The reason is that the shot noise in Eq.~(\ref{Pgg_rsd}) smooths the matter power spectrum on small scales and partially breaks the degeneracy between $a$ and $d$. A negative correlation between $d$ and $P^{\rm s}$ is then induced. Finally, the estimate of $a \approx 2$ is now compatible with expectations~\cite{Alam:2016hwk,Gil-Marin:2015sqa} and the limits on $M_{\nu}$ are considerably improved, reaching $M_{\nu}<0.22\, {\rm eV}$ at 95\%~C.L..

The addition of \clkg measurements leads to the bounds reported in the sixth and seventh row. For both the \textbf{\textbf{\textit{CMB}}}+$\boldsymbol{P_{gg}(k)}$ and \textbf{\textbf{\textit{CMB}}}+$\boldsymbol{P_{gg}(k)}$+\clkgb combinations we find a negative $d$, in agreement with the expectations from N-body simulations~\cite{Villaescusa-Navarro:2013pva,Vlah:2013lia}. The bound reported on $M_{\nu}$ for the \textbf{\textit{CMB}}+$\boldsymbol{P_{gg}(k)}$+\clkgb dataset combination ($M_{\nu}<0.19\, {\rm eV}$ at 95\%~C.L.) is the strongest available bound in the literature obtained when considering comparable datasets~\cite{Giusarma:2013pmn,Giusarma:2014zza,Palanque-Delabrouille:2015pga,
Zhen:2015yba,Gerbino:2015ixa,DiValentino:2015wba,DiValentino:2015sam,
Cuesta:2015iho,Huang:2015wrx,Giusarma:2016phn,Vagnozzi:2017ovm,Yeche:2017upn,
Couchot:2017pvz,Doux:2017tsv,Wang:2017htc,Chen:2017ayg,Upadhye:2017hdl,Salvati:2017rsn,Nunes:2017xon,Boyle:2017lzt,Zennaro:2017qnp,Sprenger:2018tdb,
Wang:2018lun,Mishra-Sharma:2018ykh,Choudhury:2018byy} and within the assumption of a $\Lambda$CDM model~\footnote{However, see	also~\cite{Emami:2017wqa,Hu:2014sea,Bellomo:2016xhl,Dirian:2017pwp,Renk:2017rzu,Peirone:2017vcq}.}. Previously, the study~\cite{Vagnozzi:2017ovm} obtained $M_{\nu}<0.30 \, {\rm eV}$ at 95\%~C.L. for the \textbf{\textit{CMB}}+$\boldsymbol{P_{gg}(k)}$ dataset with a scale-independent treatment of the bias.

\begin{figure}[h]
\centering
\includegraphics[width=8.5cm]{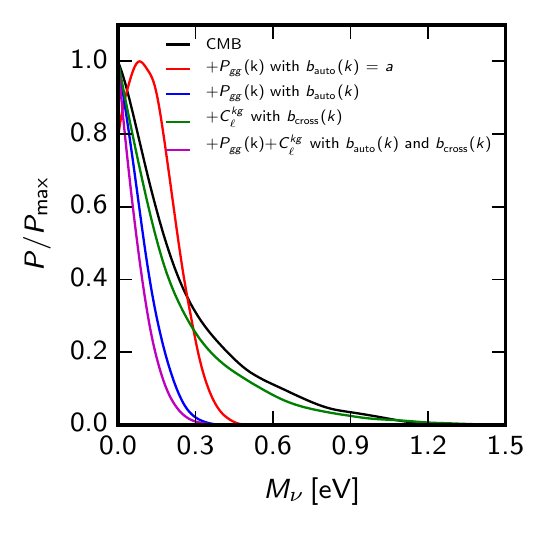}  
\caption{One-dimensional marginalized posterior for $\mnu$ obtained with the baseline \textbf{\textit{CMB}} dataset (CMB temperature and large-scale polarization anisotropy, black line), in combination with the $\boldsymbol{P_{gg}(k)}$ dataset (galaxy power spectrum from the DR12 CMASS sample, blue line), with the \clkgb dataset (CMB lensing-galaxy overdensity cross-correlation angular power spectrum, green line), and with both $\boldsymbol{P_{gg}(k)}$ and \clkgb (magenta line). We also show the posterior obtained in~\cite{Vagnozzi:2017ovm} for the \textbf{\textit{CMB}}+$\boldsymbol{P_{gg}(k)}$ dataset with a scale-independent treatment of the bias (red line).}
\label{fig:mnu}
\end{figure}

\begin{figure}[h]
\centering
\includegraphics[width=9cm]{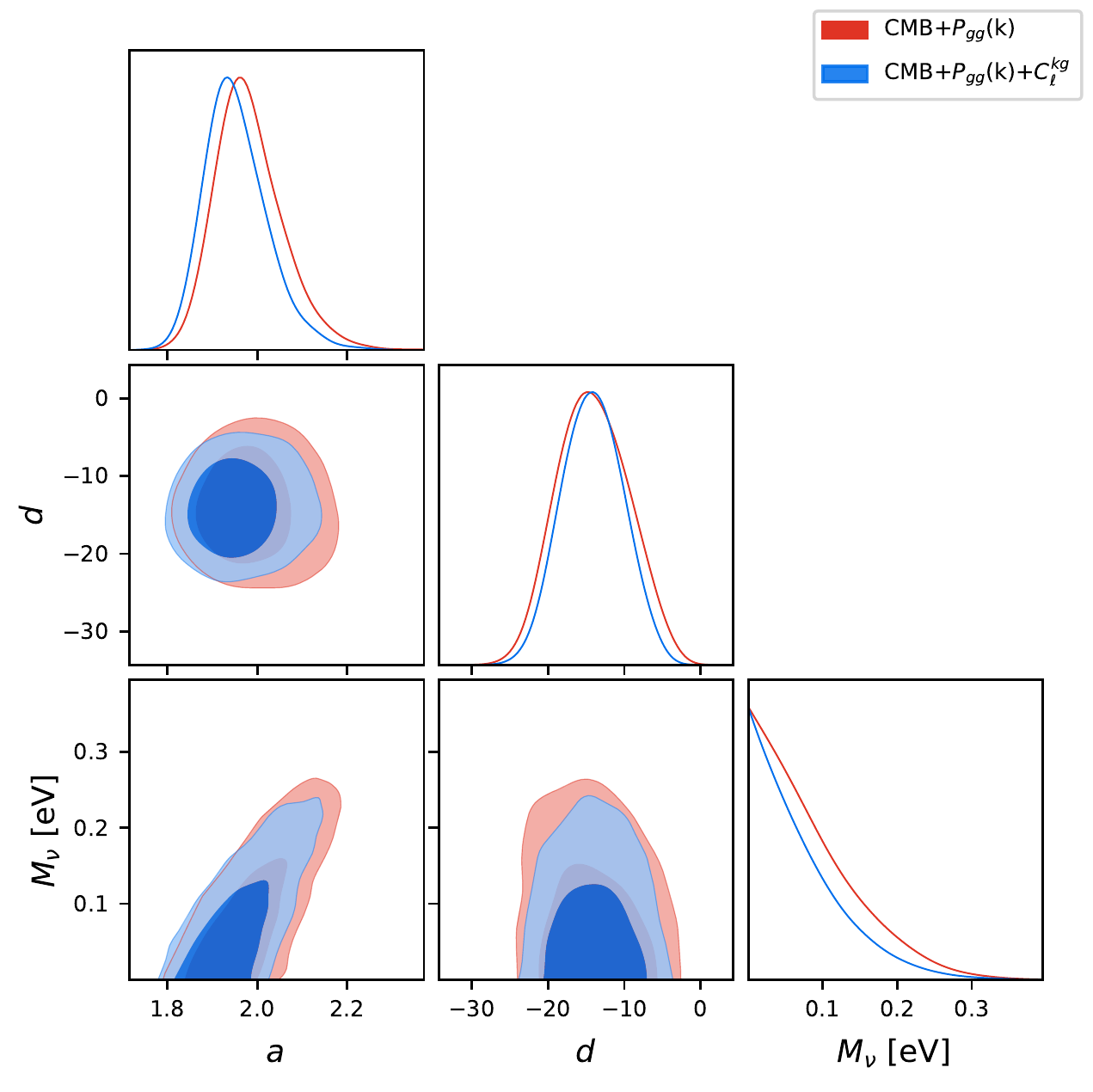}  
\caption{68\% and 95\% CL allowed regions in the combined two-dimensional planes for the parameters $\mnu$, $a$ and $d$ [the bias parameter $d$ enters the modeling of $\boldsymbol{P_{gg}(k)}$ as this is an auto-correlation measurement, see Eqs.~(\ref{galaxy}) and (\ref{biasauto})] together with their one-dimensional posterior probability distributions. We considered the combination of the \textbf{\textit{CMB}} data with the $\boldsymbol{P_{gg}(k)}$ galaxy power spectrum data (blue contours), with the further addition of the \clkgb CMB lensing-galaxy overdensity cross-correlation angular power spectrum (red contours). In order to compare these two combination of data, we do not show the parameter $c$ in the plot as it is not present in the auto-correlation parameterization [Eq.~(\ref{biasauto})].}
\label{fig:pgg_mnu_tri}
\end{figure}

The improvement in the constraints on $M_{\nu}$ can be seen in Fig.~\ref{fig:mnu}: the previous result of~\cite{Vagnozzi:2017ovm} is represented by the red curve. The small peak appearing at low values of $M_{\nu}$ has been attributed to possible systematics in the measurement, resulting in a slight suppression of small-scale power and hence a preference for higher neutrino masses. Moreover, the red curve is obtained through a scale-independent treatment of the bias [i.e. $b_{\rm auto}(k) = a$]. Thus, the results obtained using the scale-dependent expressions for $b_{\rm auto}(k)$ [Eq.~(\ref{biasauto})] and  $b_{\rm cross}(k)$ [Eq.~(\ref{biascross})] lead to a constraint on  $M_{\nu}$ which is tighter and, especially, more robust (see blue and magenta curves in Fig.~\ref{fig:mnu}). We notice that the impact of the $C_{\ell}^{\kappa g}$ dataset on improving our $M_{\nu}$ constraints is rather modest, which is best explained by the currently modest signal-to-noise of this measurement. We expect that future high signal-to-noise measurements of $C_{\ell}^{\kappa g}$, in combination with a reduction of systematics, should significantly increase the impact of this dataset, and therefore of our methodology, on constraining the cosmological parameters. Finally, triangular plots showing the joint posteriors on $a$, $d$, and $M_{\nu}$ are shown in Fig.~\ref{fig:pgg_mnu_tri}.

The bounds obtained are among the most conservative in the literature, given the bare minimum number of datasets adopted. We expect that the addition of geometrical information from BAO measurements would contribute strongly to further lowering the upper bound on $M_{\nu}$. This might open the doors towards possibly unraveling the neutrino mass hierarchy from cosmology~\cite{Huang:2015wrx,Vagnozzi:2017ovm,Allison:2015qca,Hannestad:2016fog,
Xu:2016ddc,Gerbino:2016ehw,Simpson:2017qvj,Schwetz:2017fey,Hannestad:2017ypp,
Long:2017dru,Gariazzo:2018pei,Heavens:2018adv,deSalas:2018bym}, due to parameter space volume effects. The neutrino mass bounds, and accordingly the volume effects, are actually stronger in dynamical dark energy models where $w(z)\geq-1$~\cite{Vagnozzi:2018jhn} (see also~\cite{Zhang:2015uhk,Wang:2016tsz,Zhao:2016ecj,Guo:2017hea,Zhang:2017rbg,Li:2017iur,Yang:2017amu} for related work).

\section{Conclusions}\label{sec:concl}

In this work, it is the first time that measurements of the cross-correlation between CMB lensing and galaxy overdensity maps [\clkg], and of the galaxy power spectrum [$P_{gg}(k)$], have been: a) combined and analysed in light of a well-motivated parametrization of the scale-dependent bias $b(k)$ and b) used to obtain tighter and more robust constraints on the sum of neutrino masses $M_{\nu}$. We detect scale-dependence in the bias at moderate significance, thus showing that already on linear or mildly non-linear scales ($k<0.2\, h{\rm Mpc}^{-1}$), modeling leading-order corrections to the usually assumed constant bias is important. The upper bound on $M_{\nu}$ of $0.19\, {\rm eV}$ we have determined by combining CMB data with $P_{gg}(k)$ and \clkg measurements is among the strongest and most conservative in the literature obtained with comparable datasets~\cite{Cuesta:2015iho,Huang:2015wrx,Giusarma:2016phn,Vagnozzi:2017ovm,Palanque-Delabrouille:2015pga,Alam:2016hwk}.

We expect our method to be particularly useful for future surveys, in particular for constraining cosmological parameters or models which affect small-scale clustering or the growth of structure (for example, massive neutrinos and $\sigma_8$). Moreover, our method can be extended to a tomographic analysis, using several redshift bins, allowing one to sample more modes and constrain the time-dependent suppression in the matter power spectrum due to neutrinos~\cite{Banerjee:2016suz}. Alternatively, weak lensing surveys can be used in place of CMB lensing maps~\cite{Simon:2017osp}. In order to increase the available number of modes by modeling increasingly non-linear scales, a more accurate treatment of the scale-dependent bias is necessary~\cite{Hand:2017ilm,Modi:2017wds,Seljak:2017rmr,Upadhye:2017hdl}. It will be particularly interesting to interpret CMB lensing-galaxy cross-correlation measurements within perturbation theory frameworks, for instance within convolution Lagrangian effective field theory~\cite{Modi:2017wds}. The use of such approaches will be particularly useful when cross-correlating with future galaxy surveys which will probe higher redshifts, and hence increasingly linear scales at a given wavenumber. We plan on exploring these and other issues in future work.

Finally, we expect the signal-to-noise ratio ($S/N$) for future CMB lensing-galaxy overdensity cross-correlation measurements to improve significantly. CMB-S4 like experiments in cross-correlation with future galaxy surveys should provide a $S/N$ of $\gtrsim 150$, allowing a proper modeling for the scale-dependent bias to be made. This modeling will allow a substantial recovery of information on the matter power spectrum and improve our constraints on cosmological parameters, such as $M_{\nu}$~\cite{Seljak:2008xr,Schmittfull:2017ffw}. \\

\section*{Appendix: The bias model}

In this section we discuss our choice of the bias model, Eqs.~(\ref{biascross},\ref{biasauto}), by studying the impact of using other different functional forms and quantifying to some extent the systematic error introduced adopting an incorrect model. 

As discussed in Sec.~\ref{sec:intro}, our model for the scale-dependent galaxy bias is motivated by both theory and simulations. In particular, the $k^2$ model we adopted can be derived within at least three very different theoretical approaches to understanding galaxy bias by linking the statistics of haloes to fluctuations of the primordial density field. These three extremely well-motivated and well-studied approaches, which give the same expression for the leading terms of the scale-dependent bias, are: peaks theory with Gaussian smoothing [see Eq.~(10) in~\cite{Desjacques:2010gz}], the excursion set approach [see Eq.~(50) in~\cite{Musso:2012ch}], and the effective field theory of large-scale structure\footnote{The $k^2$-correction can be understood by looking at the derivatives of $\phi$ appearing in Eqs.~(52,53) of~\cite{Senatore:2014eva}.}. A hybrid peaks theory-excursion set approach also leads to the same form for the scale-dependent bias (see Fig.~4 of~\cite{Paranjape:2012ks}).\footnote{The $k^2$-correction can also be seen in the well-known review paper~\cite{Desjacques:2016bnm}. In particular, in Eq.~(2.66), the term $b_\delta$ coincides with the standard large-scale constant bias, while the term proportional to $b_{\bigtriangledown^2\delta}$ corresponds to a $k^2$-dependent term.} Moreover, the agreement with predictions from N-body simulations (e.g.~\cite{Villaescusa-Navarro:2013pva,Vlah:2013lia}) further lend support in favour of the robustness of our choice of bias model, as being the one most justified by theory and simulations on mildly non-linear scales.

Nevertheless, several phenomenological bias models exist and have been used in literature. For instance, some reasonable choices of bias models could be those considered in Sec.~IIA of~\cite{Smith:2006ne}. These include some well-known bias forms such as the $Q$-model of Cole \textit{et al.}~\cite{Cole:2005sx}, the model of Seo \& Eisenstein~\cite{Seo:2005ys} and variants thereof~\cite{Seljak:2000jg,Schulz:2005kj,Guzik:2006bu}, the model of Huff \textit{et al.}~\cite{Huff:2006gs}, or the power law bias model of Amendola \textit{et al.}~\cite{Amendola:2015pha}. For concreteness, we have examined how the bounds would change if we used the $Q$-model of~\cite{Cole:2005sx}:
\begin{eqnarray}
b(k) = b_Q\frac{1+Qk^2}{1+1.4k} \, ,
\end{eqnarray}
where $b_Q$ and $Q$ mimic the scale dependence of the power spectrum at small scales. 

After marginalizing over $b_Q$ and $Q$, we find that also for this bias model, as for the one we used in our manuscript, the upper limit on $M_{\nu}$ is tighter than the one obtained using a scale-independent bias model. The reason is that the Monte Carlo shows a preference for values of $Q$ which result in the value of the bias decreasing as $k$ is increased (i.e. $db/dk<0$). This is exactly the same behavior we observed using our $k^2$ model, where the data prefers negative values of the $d$ bias parameter (in agreement with theoretical arguments and simulations, although at no point in the analysis have we used this information, \textit{i.e.} the prior on $d$ was large enough that the data would have been free to choose positive values of $d$ as well). In other words, galaxy power spectrum data, when interpreted using the bias models we examined, seem to prefer a bias which decreases when moving towards smaller scales: this effect can naturally be compensated by decreasing $M_{\nu}$, in order to reduce the small-scale suppression in the power spectrum caused by neutrino masses. Notice that this behavior is exactly what is expected from N-body simulations~\cite{Villaescusa-Navarro:2013pva,Vlah:2013lia}. Of course, we cannot confirm that this behavior occurs for any possible scale-dependent bias model one can think about, but the results of N-body simulations as well as our investigation of two independent bias models (the $Q$-model and the $k^2$ model we examined here) suggests that this might well be the case. A complete investigation, however, is well beyond the scope of our work. It would definitely be interesting to return to this point in more detail in the future.

\begin{figure}[!htb]
\centering
\includegraphics[width=8.5cm]{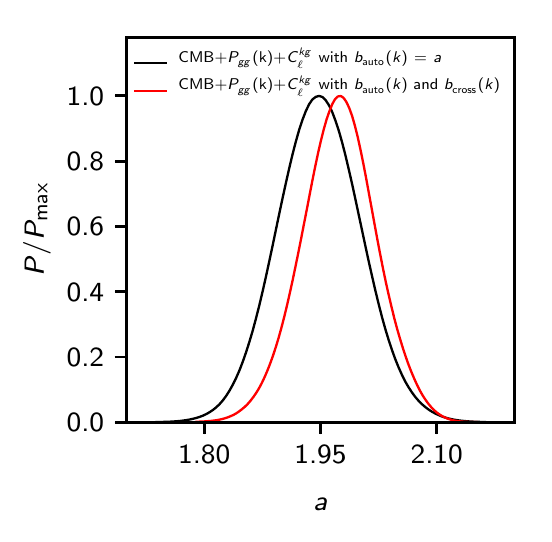}
\caption{One-dimensional marginalized posterior for $a$ (scale-independent bias parameter) obtained by combining the baseline CMB dataset, with the $P_{gg}(k)$ dataset and with the $C_{\ell}^{\kappa g}$ dataset used in this work. The red line shows the posterior obtained introducing the $k^2$-correction, while the black line illustrates the posterior obtained with a scale-independent treatment of the bias. The $k$ and $\ell$ range we choose are the same for both the cases considered.}
\label{fig:systematic}
\end{figure}

Finally, in order to somehow quantify the systematic error due to the choice of the bias model, we opted for providing a qualitative assessment by comparing the posteriors we obtain for the scale-independent bias parameter $a$, according to whether or not the $k^2$-correction is switched on (i.e., in one case we allow $c$ and $d$ to vary, and in the other case we set $c=d=0$). We plot the results in Fig.~\ref{fig:systematic}, with the red curve being the one obtained when the full scale-dependent bias model is used, whereas the black curve is obtained by considering the extreme case where we switch off the scale-dependent correction. As we can see from Fig.~\ref{fig:systematic}, the shift in the posterior of $a$ induced by introducing or not the scale-dependent correction is minimal, well below the $1\sigma$ level. From a qualitative point of view, we can expect that an incorrect bias model would lead to systematics in the recovered value of the $a$ bias parameter, which instead we find to be in agreement with the theoretical value for the galaxy sample in question ($a \sim 2$).

\begin{acknowledgments}
We are indebted to Anthony Pullen for providing the \clkg measurements and for extremely useful discussions in this respect. We thank Shadab Alam, Federico Bianchini, Emanuele Castorina, Chang Hoon Hahn, Siyu He, Alex Krolewski, Elena Massara, Patrick McDonald, Uro\v{s} Seljak, Ravi Sheth, and Martin White for useful discussions. We also thank Sebastian Baum, Alex Millar, Janina Renk, and Luca Visinelli for comments on an earlier version of the draft. This work is based on observations obtained with Planck (\href{http://www.esa.int/Planck}{\href{www.esa.int/Planck}{www.esa.int/Planck}}), an ESA science mission with instruments and contributions directly funded by ESA Member States, NASA, and Canada. We acknowledge use of the Planck Legacy Archive. We also acknowledge the use of computing facilities at NERSC and at the McWilliams Center for Cosmology. E.G. is supported by NSF grant AST1412966. S.V. and K.F. acknowledge support by the Vetenskapsr\aa det (Swedish Research Council) through contract No. 638-2013-8993 and the Oskar Klein Centre for Cosmoparticle Physics. S.H. acknowledges support by NASA-EUCLID11-0004, NSF AST1517593 and NSF AST1412966. S.F. thanks  the  Miller Institute for Basic Research in Science at the University of California, Berkeley for support. K.F. acknowledges support from DoE grant DE-SC0007859 at the University of Michigan as well as support from the Leinweber Center for Theoretical Physics.
\end{acknowledgments}

\begin{table*}
\begin{ruledtabular}
\begin{tabular}{cccccc}
Dataset	& $a$ (68\%~C.L.) & $c$ (68\%~C.L., $h^{-2}\,{\rm Mpc}^2$) & $d$ (68\%~C.L., $h^{-2}\,{\rm Mpc}^2$) & \multicolumn{2}{c}{$\mnu$ [{\rm eV}] (95\%~C.L.)}\\ 

\hline
\textit{CMB} $\equiv$ \textit{PlanckTT}+\textit{lowP} &  & & & $< 0.72$ & [$<0.77$] \\
\textit{CMB}+\clkg & $1.45 \pm 0.19$ & $2.59 \pm 1.22$ & &  \textbf{0.06}  & \\
                             & $1.50 \pm 0.21$ & $2.97 \pm 1.42$ & & $< 0.72$ & [$<0.77$] \\
\textit{CMB}+$P_{gg}$(k) & $1.97\pm0.05$ &   & $-13.76\pm4.61$& \textbf{0.06} & \\
					& $1.98\pm0.08$ &  & $-14.03\pm4.68$ & $< 0.22$ &[$<0.24$] \\
\textit{CMB}+$P_{ gg}$(k)+\clkg & $1.95 \pm 0.05$ &   $0.45\pm0.87$& $-13.90 \pm 4.17$  & \textbf{0.06}  &\\
  & $1.95 \pm 0.07$ & $0.48\pm0.90$ & $-14.13 \pm 4.02$ & $< 0.19$ & [$<0.22$] \\
\end{tabular}
\caption{Constraints on the bias parameters $a$, $c$, and $d$, as well as the sum of the three active neutrino masses $\mnu$. The bounds on $\mnu$ not in square brackets have been obtained imposing a lower bound of $\mnu > 0\, {\rm eV}$, i.e. only making use of cosmological data, whereas the ones in square brackets have been obtained imposing the lower bound set by neutrino oscillations of $\mnu>0.06\, {\rm eV}$. The \textbf{\textit{CMB}} dataset denotes measurements of the CMB temperature and large-scale polarization anisotropy from the Planck satellite 2015 data release. Measurements of the angular cross-power spectrum between CMB lensing convergence maps from the Planck 2015 data release and galaxies from BOSS DR11 CMASS sample [\clkgb], as well as the galaxy power spectrum measured from BOSS DR12 CMASS sample [$\boldsymbol{P_{gg}(k)}$], are then added. Rows featuring the symbol \textbf{0.06} were obtained fixing the sum of the neutrino masses $M_{\nu}$ to the minimum value allowed by oscillation data, $0.06\, {\rm eV}$.}
\label{tab:tab1}
\end{ruledtabular}
\end{table*}

\end{document}